\documentstyle[11pt,a4wide,epsf]{article}

\newcommand{\pspicture}[1]{
\centerline{\setlength\epsfxsize{9.2cm}\epsfbox{#1}}}

\def\NPB{{\em Nucl. Phys.} B}

\def\PLB{{\em Phys. Lett.}  B}
\def\PRL{{\em Phys. Rev. Lett.}}
\def\PRD{{\em Phys. Rev.} D}

\newcommand{\mh}{m_H}
\newcommand{\GeV}{\mbox{GeV}}
\newcommand{\order}[1]{{\cal O}(#1)}
\newcommand{\sutwo}{SU(2)}

\begin{document}
\pagestyle{plain}
\begin{flushright}
DAMTP-98-12\\
UTCCP-P-37\\
\end{flushright}
\vskip 20pt

\begin{center}
{\large \bf The  Chern-Simons number as an order parameter : 
classical sphaleron transitions for $SU(2)$-Higgs
field theories 
for $\mh  \approx  120 \, \GeV$ } \\[10pt] 
H.P.~Shanahan\footnote{shanahan@rccp.tsukuba.ac.jp},\\[10pt]
{\sl Center for Computational Physics, University of Tsukuba,
Tsukuba, Ibaraki 305, Japan} \\[10pt] 
A.C.~Davis\footnote{A.C.Davis@damtp.cam.ac.uk},\\[10pt]
{\sl DAMTP, University of Cambridge, Cambridge CB3 9EW, England}
\end{center}

\begin{abstract}
The classical transitions between  topologically distinct 
vacua in a $SU(2)$--Higgs model,  using  
a Higgs field of mass approximately  $120 \, \GeV$, is examined 
to probe the crossover region between the symmetric and broken phase.
Assuming the Higgs mass is constant, we find 
the width of this crossover region is approximately 20\% of the 
average temperature. 
We suggest that this observable is a better parameter
to explore this region of phase space than 
equal-time correlation
functions.
\end{abstract}

\section{Introduction}
There is evidence from a number of numerical simulations in 3 and 4 
dimensions \cite{DESY-static-therm,Finn-static-therm, Aoki}
that the first order phase transition from the symmetric to broken phase 
of the electroweak sector of the Standard Model
ends for a Higgs mass of approximately  $80 \, \GeV$. 
For larger Higgs masses, there is analytic crossover or very possibly 
a transition where higher order derivatives of the free energy 
experience a discontinuity (these are sometimes referred to as 
``third or higher order'' phase transitions, whether numerical studies 
could ever distinguish between these two scenarios remains to be seen). 
The preliminary lower bound  of $77 \, \GeV$ for the Higgs  
mass \cite{LEP-Higgs}
indicates that this latter  region of phase space, 
where the behaviour of equal time correlation functions is much smoother, 
is physically far more likely.

This endpoint, and the fact that the first order phase transition
is very weak for smaller Higgs masses \cite{hein} indicates that 
baryogenesis cannot occur  via  this mechanism in the Standard 
Model (SM). 
As a result, a number of other suggestions  have
been made to get around this. Many of these proposals make
non-trivial statements about the early Universe.
While the qualitative picture 
of supersymmetric models is no different from the SM,
it is hoped that the extra fields will adjust the position
and strength 
of the phase transition sufficiently that larger Higgs masses
are feasible \cite{susy}. 
Other models introduce extensions to the SM, including the formation 
of topological defects to provide the out of equilibrium condition
\cite{brandenberger}, or change the 
expected rate of expansion of the Universe at the EW epoch \cite{Joyce} 
so that a phase transition is unnecessary and 
only sufficiently rapid change is sufficient.

It is therefore of interest to explore the baryon number
violating properties in this region of parameter space. 
A study to see if defect formation can occur in this region would
also  be of interest. In this paper we shall focus on the first
point.
Here we attempt to calculate the rate of diffusion for the Chern-Simons 
number defined as 
\begin{equation}
N_{CS}(t)  - N_{CS}(0) = 
\frac{1}{32 \pi^2} \int_0^t dx^0 \int d^3x 
\epsilon^{\mu\nu\rho\lambda} {\bf F}_{\mu\nu}. {\bf F}_{\rho\lambda} \;\; ,
\end{equation}
which labels the change in the baryon number as a function of 
time (for a recent review see \cite{thooft}).
Such an unequal-time correlation function 
is extremely difficult to calculate
in a full quantum field theory, discretised on a lattice, 
as the introduction of a 
density matrix operator ensures that the exponential weight in a
Monte-Carlo simulation will be complex and will fluctuate enormously.
However, noting 
that on dimensionful grounds the typical size of the sphaleron
is $\order{1/gT} >> T^{-1}$, where $T$ is the temperature, 
a classical interpretation
of the problem seems to be a reasonable approximation. The accuracy 
of such an approach is of some debate but has been successfully  
utilised by an number of different groups 
for pure $SU(2)$ gauge theories
and $SU(2)$--Higgs systems with $\mh  \le 80 \, \GeV$ 
\cite{ambjorn,smit,moore}.
Here, we choose a Higgs mass of approximately
$120 \, \GeV$ and have determined the diffusion rate
for a range of $aT$, where $a$ is the lattice spacing, and two volumes.
This paper proceeds as follows; 
in section 2 we outline briefly the method of calculation, describing
the discretised Hamiltonian and the choice of bare lattice parameters.
In section 3 we present an analysis of the results and determine their
behaviour in terms of physical observables. Finally we draw some 
conclusions on the applications of this approach.
 
\section{Computational Details}
We employ the methods used in \cite{ambjorn,smit,moore,moore-slave} 
to evaluate
the {\it classical} diffusion rate of the Chern Simons number. 
The original 4-dimensional $\sutwo$-Higgs lagrangian is dimensionally 
reduced by the high-temperature Matsubara formalism to a surprisingly
simple 3-dimensional lagrangian, whose couplings are related
to the original lagrangian's coupling by perturbation 
theory \cite{shaposhnikov}.  
To measure a quasi-equilibrium quantity such as the above
diffusion rate, one introduces conjugate momenta fields 
which allow the fields to evolve in a Hamiltonian
formalism, while staying in thermal equilibrium.
The $\sutwo$ scalar and gauge fields are represented in a discrete 
form by the fields $\Phi_{\bf x}$ and $U_{i,{\bf x}}$ respectively,
where $\bf x$ is the site index.
The conjugate momenta fields are $\pi_{\bf x}$ and $E_{i,{\bf x}}$.
The following discretised Hamiltonian was employed 
\begin{eqnarray}
H \hskip-10pt &=& \hskip-10pt
 \beta\left[ - \sum_{\bf x}\left(1 - \frac{1}{2}Tr U_{\Box,{\bf x}}\right) -
\frac{1}{2} \sum_{{\bf x},i} ( \Delta_i \Phi )_{\bf x}^\dagger 
( \Delta_i \Phi )_{\bf x} \right. 
  \label{eqn:hamiltonian} \nonumber  \\ 
 &-& \hskip-10pt\left.   \sum_{\bf x}\left( \frac{M^2_{H0}}{2} 
\Phi_{\bf x}^\dagger \Phi_{\bf x} +
\frac{\lambda_L}{4} 
(\Phi_{\bf x}^\dagger \Phi_{\bf x})^2 \right) 
+ \frac{1}{(\Delta t)^2}\left( z_E 
\sum_{i,{\bf x}} E_{i,{\bf x}} E_{i,{\bf x}} + \frac{z_\pi}{2} 
\sum_{\bf x} 
\pi_{\bf x}^\dagger \pi_{\bf x}\right) \right] \;\; ,
 \end{eqnarray}
where 
\begin{equation}
\left(\Delta_i \Phi\right)_{\bf x} = 
\Phi_{\bf x + \hat{i}} U_{i,{\bf x}} - \Phi_{\bf x} \;\; ,
\end{equation}
and $U_{\Box,{\bf x}}$ is the elementary plaquette constructed from the
gauge fields surrounding that point.
The parameter $\Delta t$ controls the time step size and was set to 
$0.05$. The parameters $z_E$ and $z_\pi$ represent the effect of 
the renormalisation of the momentum operators by the dimensional 
reduction and is assumed to take the form $1 + \order{g^2}$. 
The validity of this assumption (or indeed whether higher dimensional 
operators should be included) is still a topic of discussion. 
No estimate of the one-loop corrections to $z_E$ and $z_\pi$ have been
made and we have assumed  them to be 1.
The remaining parameters $\beta$, $M^2_{H0}$ and $\lambda_L$ control
the  physical parameters of the system;  the lattice spacing 
$a$, the Higgs mass $m_H$ and the temperature $T$. At this level
of approximation, it is assumed that the  couplings $g^2$
and $\lambda$ (for the original  4-dimensional lagrangian)
do not run and that many of the standard tree-level results can be applied.
Noting that 
\begin{equation}
\beta = \frac{4}{g^2 a T} \;\; , 
\end{equation}
and that at tree-level
\begin{equation}
\lambda_L \approx 8 \frac{\lambda}{g^2}   \approx \frac{m_H^2}{m_W^2} \;\; , 
\end{equation}
it is clear that one can fix $\mh$  and $aT$ simply from
the $\beta$ and $\lambda_L$.
The relationship that  the parameter 
$M_{H0}^2$ has to $\mh$, $a$ and $T$ is more complicated.
One could vary $M_{H0}^2$ and $\beta$ so that $a$ is fixed. 
However, for this 
paper we shall take a more simple route by fixing $\lambda_L$
and $M_{H0}^2$ and varying $\beta$. 
The ratio $T/\mh$ as a function of $\beta$ 
can be determined from a perturbative relationship \cite{shaposhnikov}. 

For this paper, $\lambda_L$ was set to $2.25$  and $M_{H0}^2$ to $-0.596$
using two lattice sizes of $24^3$ and $18^3$. The coupling $\beta$ was
varied from 6.8 to 7.8 in steps of 0.1. For each $\beta$, 8 to 16 
configurations were thermalised and then evolved using a leapfrog algorithm 
for time lengths greater than 6000. The Chern-Simons number was measured using
the slave-field technique devised by Moore and Turok \cite{moore-slave} at
intervals of  5 time units. The simulation parameters are listed in
table (\ref{tab:params}). The code used for thermalisation and
evolution  was developed by Guy  Moore and Neil Turok and was run on 
a Silicon Graphics Origin 2000 made available by the 
U.K. Computational Cosmology
Consortium (UKCCC) and an Hitachi SR2201 at the High Performance
Computing Facility (HPCF) at the University of Cambridge.

\section{Results}
\begin{figure}[t]

\pspicture{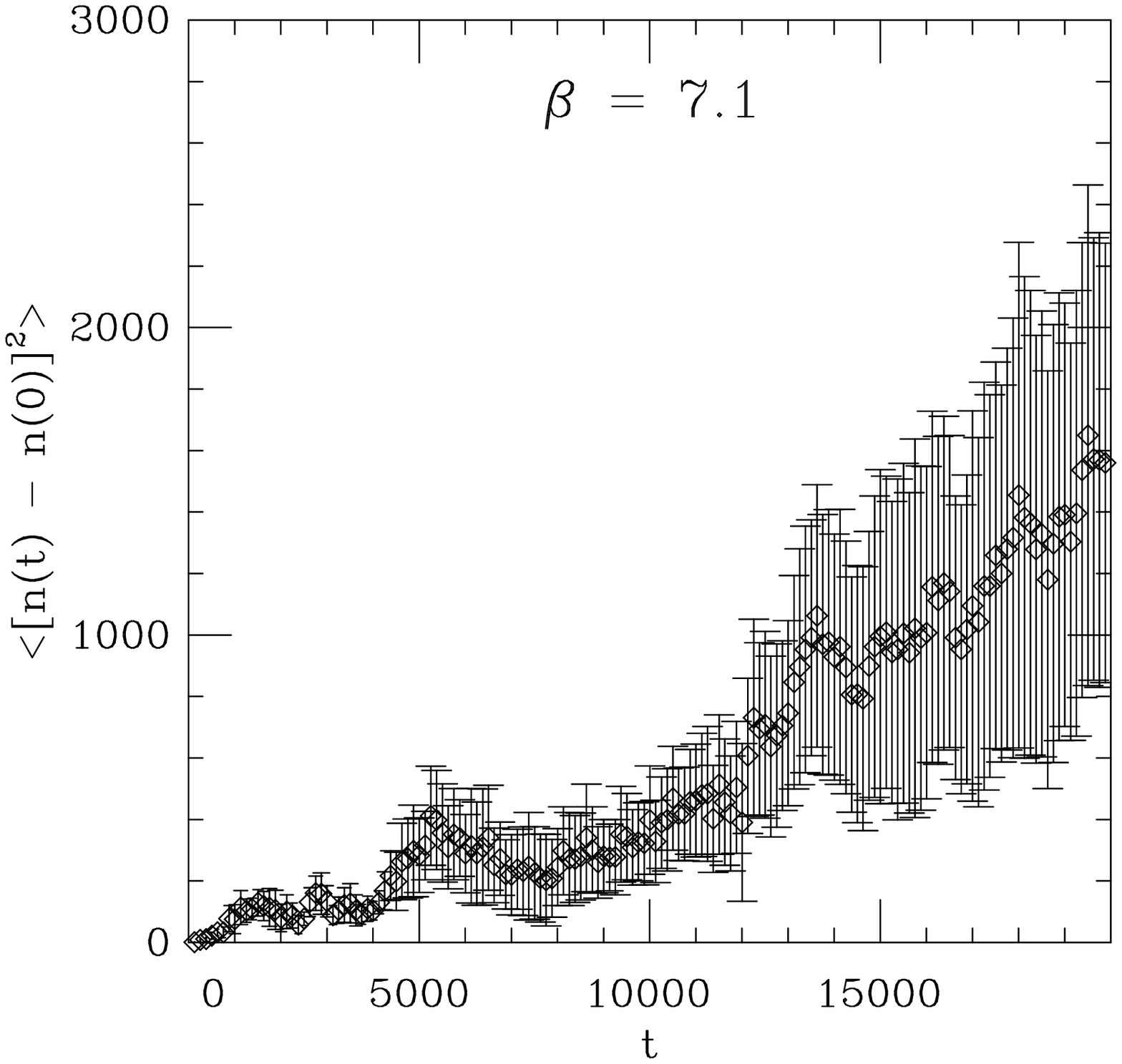}
\pspicture{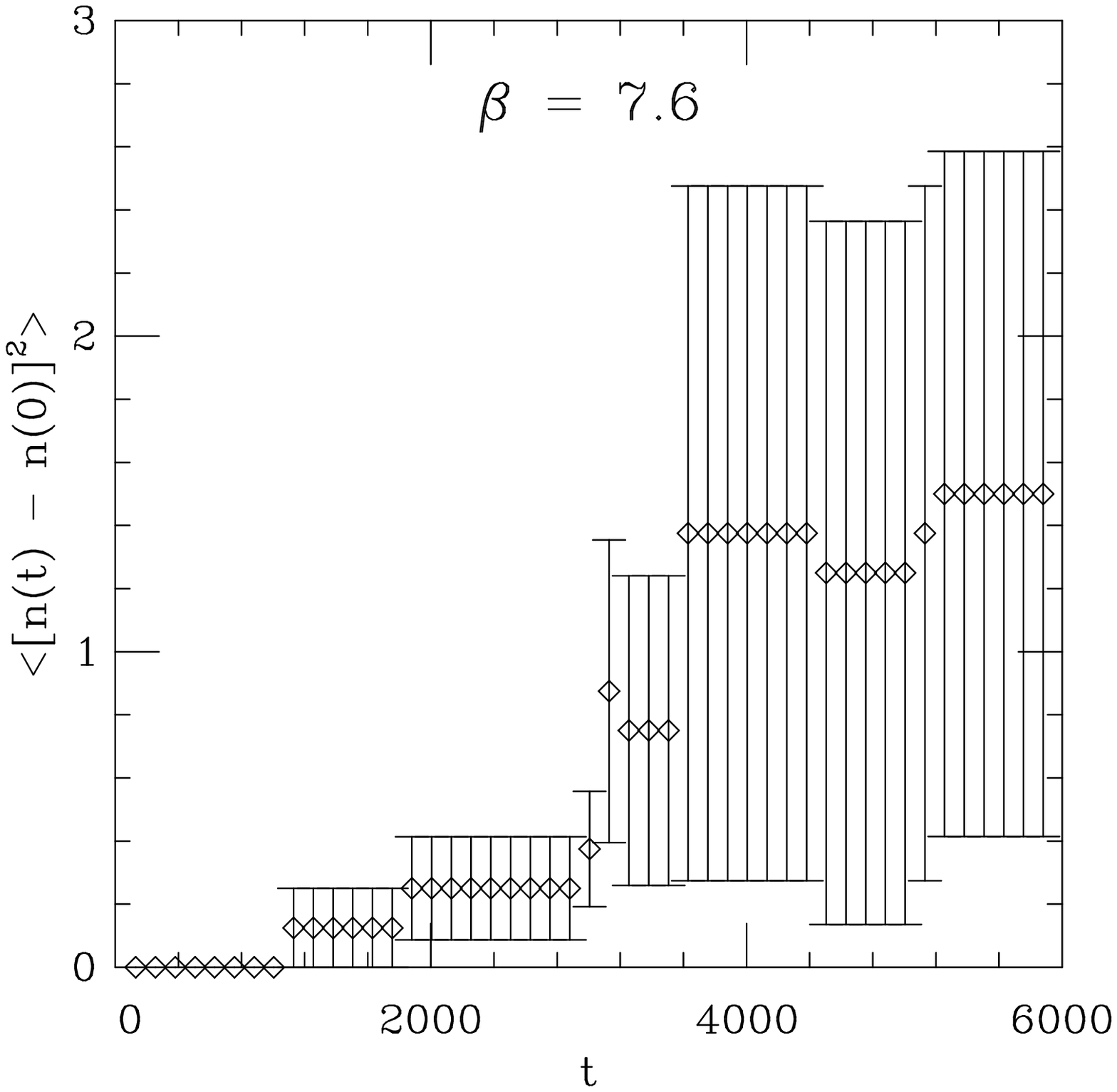}

  \caption{Diffusion rate of winding number at $\beta=7.1$ and $\beta=7.6$. 
Note the difference in the vertical scales.}
\label{fig:random_walk}
\end{figure}

\begin{figure}[t]
\pspicture{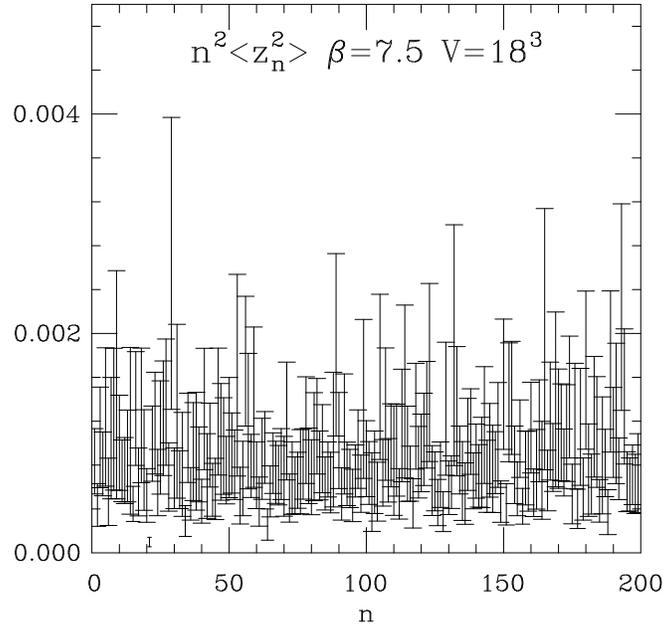}
\caption{Cosine transform coefficients for a typical parameter set.}
\label{fig:cosines}
\end{figure}

\begin{figure}[t]
\pspicture{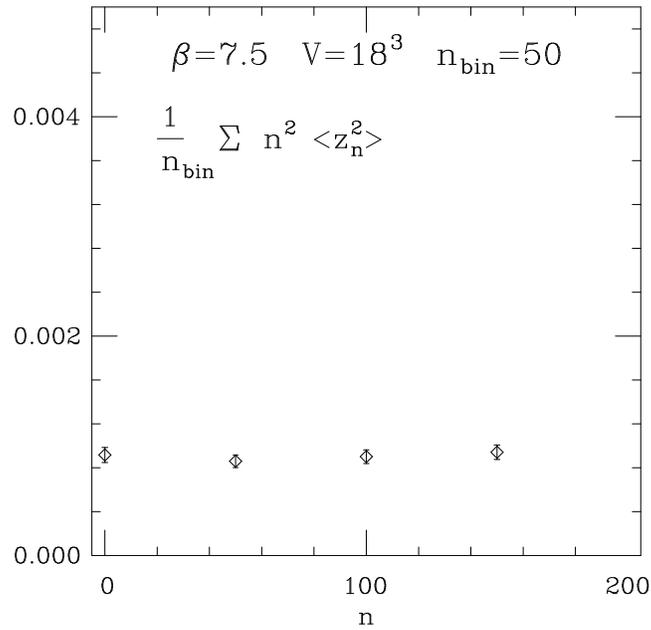}
  \caption{Cosine transform coefficients binned into groups of 50, 
using the same parameter set as in figure (\ref{fig:cosines})}
\label{fig:bin_cosine}
\end{figure}

Assuming the behaviour of $N_{CS}$ is diffusive, then 
\begin{equation}
\lim_{t \rightarrow \infty} <(N_{CS}(t) - N_{CS}(0))^2> =
\Gamma t \;\; . 
\end{equation}
Examples of this expectation value as a function of time
are shown in figure (\ref{fig:random_walk}). To isolate
$\Gamma$ the  cosine transform technique, 
outlined in  \cite{moore-slave}, was used. 
This involves integrating the data with a cosine weighting over
the trajectory length. For a finite
time spacing $\Delta t$ and a trajectory length of $t_f + \Delta t$
 this corresponds to evaluating
\begin{equation}
z^I_n  = \frac{1}{t_f + \Delta_t} 
\left[ \left(\sum_{t=\Delta_t}^{t_f} 
N^I_{CS}(t) \cos{\frac{n \pi t}{t_f}} \right)
- \frac{1}{2} \left( N^I_{CS}(0) + N^I_{CS}(t_f) \cos{n\pi} \right) \right] 
\; \;,
\end{equation}
where the index $I$ corresponds to an individual configuration.
Averages over configurations are denoted with $<\; >$.
An example of the resulting coefficients are shown in figure 
(\ref{fig:cosines}). 
It is possible to demonstrate that
\begin{equation}
n^2 <z^2_n> = \frac{\Gamma t_f}{2 \pi^2} + \zeta n^2 \;\; ,  
\end{equation}
where $\zeta$ is a white noise term, although a functional fit to 
this form was not used.
The coefficients $n^2 <z_n^2>$ were averaged in
bins of 50 which considerably reduced the resulting error. An example
of this is shown in in figure (\ref{fig:bin_cosine}).
The final central value
was determined from the central value of the binned data of order
50 to 99, as higher orders will be affected by ultra-violet contributions
while smaller coefficients may be affected by the finite length of
the trajectory. 
If the other binned coefficients of 1-49 or 100-149
varied by more than a standard deviation from this,
the difference was included as
a systematic error.

In order to express the results in terms of  a dimensionless intensive 
quantity $\overline{\kappa}$,  the diffusion rate in the continuum
$\Gamma^{cont}$
is usually expressed as 
\begin{equation}
\frac{\Gamma^{cont}}{V}  = \overline{\kappa} (\alpha_W T)^4 \; \; ,
\end{equation}
(although it remains unclear if $\overline{\kappa}$ has some residual 
dependence on $\alpha_W$ \cite{arnold}).
Equivalently, for the lattice result
\begin{equation}
\Gamma = V (\pi \beta)^{-4} \kappa \; \; .
\end{equation}
The results for  $\kappa$  are listed
in table (\ref{tab:params}).

\section{Conclusions}
In this letter, we have calculated the diffusion rate of the Chern Simons
number for an $\sutwo$--Higgs field theory at approximately fixed 
Higgs mass 
for a range of temperatures. 
We see that the rate is independent of volume and varies
from a constant non-zero rate, in the symmetric phase to a zero rate
in the broken phase (the use of the word ``phase'' here is perhaps 
slightly vague and should be redefined in this case by 
the region of parameter space with properties similar to the 
other region of parameter space where a phase transition occurs).
The ratio  $\mh/T$ can be determined 
from perturbative matching of the 3 and 4 dimensional coefficients
as described in \cite{shaposhnikov} and \cite{smit}. 
The change from a zero to non-zero rate occurs 
for a variation in $\mh/T$ of approximately
20\%. This is broader than the case when $\mh \approx 80 \, \GeV$,
which was approximately 10\% \cite{smit}.  
Assuming the Higgs mass is fixed at $120 \; \GeV$ we see that the
mean crossover  temperature ($\sim380 \, \GeV$) 
is larger than is expected from the   
study of equal-time correlation functions ($\sim210 \, \GeV$) 
\cite{Finn-static-therm}.  
A more precise study in parameter space is certainly possible,
where the lattice spacing is kept constant and only the temperature
varied. This would involve calculating the Higgs and $W$ mass  
for each possible configuration of parameters to determine
the lattice spacing. 
However, as it unclear how accurate the classical
approximation is, such an exhaustive approach may be fruitless.

Nonetheless, the Chern-Simons diffusion rate is an excellent
observable for studying the complicated crossover region of such
field theories. Equal-time correlation functions provide very
clear signals in the region where a first-order phase transition
exists.  However in the region of large Higgs mass, their behaviour is
non-singular and the distinction between analytic crossover and possible
discontinuities becomes quite difficult. 
The Chern-Simons diffusion rate has behaviour that is clearer and 
independent of volume  which defines two separate regions. 

One area this study may well be of use is in the area of defect 
formation for large-structure formation in the early Universe.
It is not clear whether defects can form in this crossover region.
In the future we plan to study this.

\section*{Acknowledgements}
The code for thermalising and evolving the configurations 
was written by Guy Moore and Neil Turok and we express
our gratitude for their generosity in letting us use this code.
We express our thanks to the UKCCC and HPCF for their support
while running this code. We thank Ian Drummond, Arthur Hebecker and
Ron Horgan for useful discussions.
HPS is supported by the Leverhulme trust and the JSPS. This work is 
supported in part by PPARC.

\begin{table}[t]
\vspace{0.4cm}
\begin{center}
\begin{tabular}{|c|c|c|c|c|c|}
\hline
$\beta$ & $\frac{T}{m_H}$ &  $N_s^3$ & \#configurations & \#time steps & $\kappa$ \\ \hline
6.8 & 3.85 &  $24^3$ & 16 & 7000 & 0.973 $\pm$ 0.057 $\pm$ 0.161 \\
6.9 & 3.67 & $24^3$ & 16 & 7000 & 0.984 $\pm$ 0.061 $\pm$ 0.107 \\
7.0 & 3.50 & $24^3$ & 16 & 10000 & 1.013 $\pm$ 0.083 $\pm$ 0.187 \\
7.1 & 3.35 & $24^3$ & 8 & 20000 & 0.770 $\pm$ 0.059 \\
7.2 & 3.21 & $24^3$ & 8 & 10000 & 0.388 $\pm$ 0.034 $\pm$ 0.038\\
7.3 & 3.08 & $24^3$ & 8 & 10000 & 0.092 $\pm$ 0.008 \\
7.4 & 2.96 & $24^3$ & 8 & 6000 & 0.018 $\pm$ 0.002 $\pm$ 0.003\\
7.5 & 2.85 & $24^3$ & 8 & 9000 & 0.00787 $\pm$ 0.00068 \\
7.6 & 2.75 & $24^3$ & 16 & 6000 & 0.00088 $\pm$ 0.00006 \\
7.7 & 2.66 & $24^3$ & 8 & 6000 & 0.00130 $\pm$  0.00016\\
7.8 & 2.57 & $24^3$ & 8 & 9000 & 0.00055 $\pm$  0.00005\\  
\hline 
6.8 & 3.85 & $16^3$ & 16 & 10000 & 1.021 $\pm$ 0.057 $\pm$ 0.145 \\
6.9 & 3.67 & $16^3$ & 8 & 11500 & 1.058 $\pm$ 0.061 $\pm$ 0.163 \\
7.0 & 3.50 & $16^3$ & 16 & 11000 & 0.934 $\pm$ 0.058  \\
7.1 & 3.35 & $16^3$ & 16 & 7500 & 0.754  $\pm$ 0.044  $\pm$ 0.078 \\
7.2 & 3.21 & $16^3$ & 16 & 11500 & 0.276 $\pm$ 0.017 \\
7.3 & 3.08 & $16^3$ & 15 & 6000 & 0.0700 $\pm$ 0.0061 $\pm$ 0.0095 \\
7.4 & 2.96 & $16^3$ & 16 & 15500 & 0.0179 $\pm$ 0.0011 \\
7.5 & 2.85 & $16^3$ & 16 & 12500 & 0.00454 $\pm$ 0.00031 \\

 \hline
\end{tabular}
\end{center}
\caption{Numbers of configurations and the time steps iterated
forward for each $\beta$ \label{tab:sweeps}. The second error
quoted for $\kappa$ is the difference between the first and second bins
of 50 coefficients when the difference was greater than 1 standard deviation.}
\label{tab:params}
\end{table}

\begin{figure}[t]

\pspicture{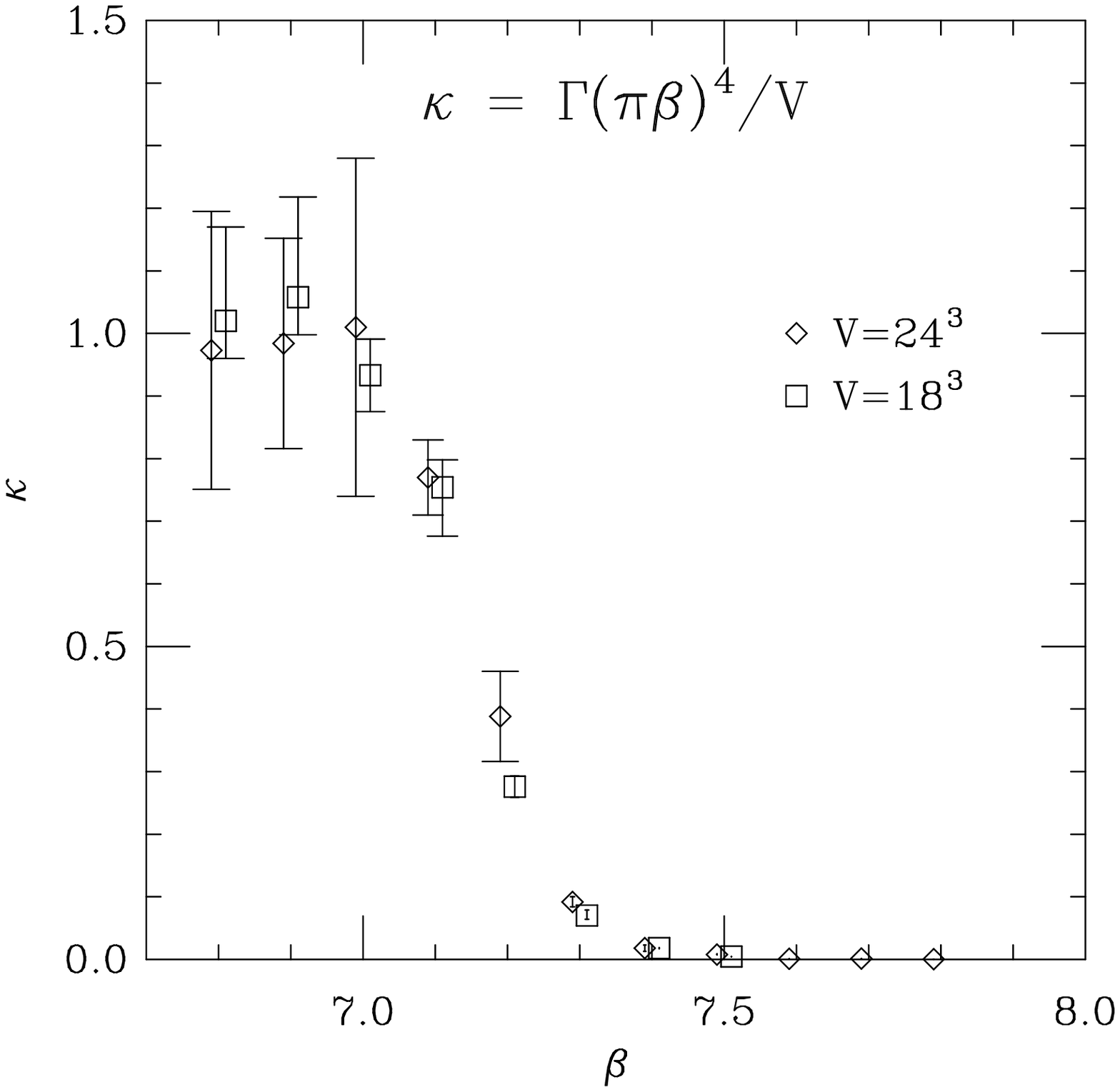}

  \caption{Resultant $\kappa$'s for both volumes as a function of 
$\beta$. The results for different volumes are displaced slightly 
for clarity.}
\label{fig:kappa_final}
\end{figure}


\begin{thebibliography}{99}
\bibitem{DESY-static-therm} 
F.~Csikor {\it et al},
{\em Nucl. Phys. (Proc Suppl.)}
{ \bf 53}, 612-614, (1997).
\bibitem{Finn-static-therm}
K.~Kajantie {\it et al}, 
\PRL {\bf 77}, 2287-2890, (1997).
\bibitem{Aoki}
Y.~Aoki,
\PRD {\bf 56}, 3860-3865, (1997).
\bibitem{LEP-Higgs}
W.~Murray, International Europhysics  conference on High
Energy Physics, Jerusalem, Israel,
1997 (Springer Verlag). 
http://www.cern.ch/hep97/abstract/tpa13.htm
\bibitem{hein}
K.~Kajantie {\it et al.},
\NPB {\bf 466}, 189, (1996).
\bibitem{susy}
M.~Laine,
\NPB {\bf 481}, 43-84, (1996).
\bibitem{brandenberger}
R.~Brandenberger, A.C.~Davis and M.~Trodden,
\PLB {\bf335}, 123-130 (1994);
R.~Brandenberger, A.C.~Davis, T.~Prokopec and M.~Trodden,
\PRD {\bf 53}, 4257-4266 (1996).
\bibitem{Joyce}
M.~Joyce, 
E\"otv\"os conference in science: Strong and Electroweak matter 
(SEWM 97), Eger, Hungary, 1997, hep-ph/99709321.
\bibitem{thooft}
V.A.~Rubakov and M.E.~Shaposhnikov,
{\em Usp.Fiz.Nauk}  166  493-537 (1996); 
{\em Phys.Usp. } 39  461-502 (1996),
hep-ph/9603208.
\bibitem{ambjorn}
J.~Ambjorn and A.~Krasnitz,
\PLB {\bf 362}, 97-104,(1995).
\bibitem{smit}
W.~Tang and J.~Smit,
\NPB {\bf 482}, 265-285,
\bibitem{moore}
G.~Moore and N.~Turok,
\PRD {\bf 55}, 6538-6560, (1997). 
\bibitem{moore-slave}
G.~Moore and N.~Turok,
\PRD {\bf 56}, 6533-6546, (1997). 
\bibitem{shaposhnikov}
K.~Kajantie {\it et al.} 
\NPB {\bf 458}, 90, (1996).
\bibitem{arnold}
P.~Arnold,
\PRD {\bf 55}, 6264-6273, (1997).
\end{thebibliography}
\end{document}